\documentclass[sigconf]{acmart}
\usepackage{algorithmic} 
\usepackage{graphicx}
\usepackage{textcomp}
\usepackage{xcolor}
\usepackage{url}
\usepackage{todonotes}
\usepackage{listings}
\usepackage{float}
\usepackage{hyperref}

\AtBeginDocument{%
  \providecommand\BibTeX{{%
    \normalfont B\kern-0.5em{\scshape i\kern-0.25em b}\kern-0.8em\TeX}}}

\renewcommand\footnotetextcopyrightpermission[1]{} % removes footnote with conference information in first column
\begin{document}

\title{A Study of Single Statement Bugs Involving Dynamic Language Features}

% \author{ Anonymous}
% \author{Li Sui, Shawn Rasheed and Amjed Tahir}
% \affiliation{%
%   \institution{Massey University}
%   \city{Palmerston North}
%   \country{New Zealand}}

\author{Li Sui, Shawn Rasheed, Amjed Tahir}
\affiliation{%
  \institution{Massey University}
  \city{Palmerston North}
  \country{New Zealand}}

%\email{leesui0207@gmail.com,{S.Rasheed,A.Tahir}@massey.ac.nz}

% \author{Shawn Rasheed}
% \affiliation{%
%   \institution{Massey University}
%   \city{Palmerston North}
%   \country{New Zealand}}
% \email{S.Rasheed@massey.ac.nz}

% \author{Amjed Tahir}
% \affiliation{%
%   \institution{Massey University}
%   \city{Palmerston North}
%   \country{New Zealand}}
% \email{A.Tahir@massey.ac.nz}

\author{Jens Dietrich}
\affiliation{%
  \institution{Victoria University of Wellington}
  \city{Wellington}
  \country{New Zealand}}
%\email{Jens.Dietrich@vuw.ac.nz}

%\thispagestyle{plain}
%\pagestyle{plain}

\begin{abstract}
Dynamic language features are widely available in programming languages to implement functionality that can adapt to multiple usage contexts, enabling reuse. Functionality such as data binding, object-relational mapping and user interface builders can be heavily dependent on these features. However, their use has risks and downsides as they affect the soundness of static analyses and techniques that rely on such analyses (such as bug detection and automated program repair). They can also make software more error-prone due to potential difficulties in understanding reflective code, loss of compile-time safety and incorrect API usage. 
In this paper, we set out to quantify some of the effects of using dynamic language features in Java programs – that is, the error-proneness of using those features with respect to a particular type of bug known as single statement bugs.  %we want to raise a research question: how frequent that dynamic language features can appear in single statement bugs in Java programs? W
By mining 2,024 GitHub projects, we found 139 single statement bug instances (falling under 10 different bug patterns), with the highest number of bugs belonging to three specific patterns: \textit{Wrong Function Name}, \textit{Same Function More Args} and \textit{Change Identifier Used}. These results can help practitioners to quantify the risk of using dynamic techniques over alternatives (such as code generation). 
We hope this classification raises attention on choosing dynamic APIs that are likely to be error-prone, and provides developers a better understanding when designing bug detection tools for such feature. 

%with the highest number of bugs belonging to three specific patterns: \textit{Wrong Function Name}, \textit{Same Function More Args} and \textit{Change Identifier Used}. 

\end{abstract}

\maketitle

% \ccsdesc[500]{Empirical software engineering}
% \ccsdesc{Mining software repository}
% \ccsdesc[300]{Program analysis}
% \ccsdesc{Java program language}
% \ccsdesc{Java reflection}
% \ccsdesc[100]{Bug detection}

\section{Introduction}
Dynamic language features are widely available in modern programming languages and commonly used in code. In Java, features like reflection and dynamic proxies, are used to implement generic components that can be used in different contexts \cite{dietrich2017xcorpus}. Landman et al. \cite{landman2017challenges} report that 78\% of Java projects they studied contain reflective calls. However, there are challenges in modelling and analysing programs and finding bugs that involve the use of dynamic language features \cite{livshits2015defense,dietrich2017construction}. %It can also be difficult to repair such bugs as automated repair may depend on similar static analyses.
%Static analysers are known to be unsound in the presence of dynamic language features \cite{livshits2015defense}.
%that are hard to analyse with static analysis. 
Recent studies on the analysis of dynamic features have focused on improving the recall of static analysers, whilst maintaining their precision \cite{livshits2015defense,sui2020recall}. Challenges in analysing these features may affect software maintenance activities such as bug detection, automated refactoring and program repair.

The question is what the propensity is for the use of  dynamic features to cause bugs. Dynamic features can be error-prone due to erroneous interpretation of APIs, difficulty in comprehending reflective code, loss of compile-time safety and incorrect API usage. However, to the best of our knowledge, no studies on bugs related to the use of these features in Java application code have been conducted. There was a recent study on bugs involving the use of dynamic features in Python \cite{chen2018study}.% \todo[inline]{these bugs were identified in JVM's, not application code, which is what we say earlier}
and some studies on reflection specific bugs in JVM code \cite{pontes2019}. 

In this work, we attempt to answer the question: \textit{``how frequently are dynamic language features the cause of single statement bugs in Java programs?''} Single statement bugs are the class of bugs which can be fixed with a change to a single statement. This simplicity allows easy identification whether bugs are related to the use of specific APIs (e.g., reflection).
%we attempt to understand the nature of dynamic feature related single statement bugs and opportunities to leverage this knowledge in bug detection and program repair. 
To answer this question, we extracted bug instances that involve references to dynamic language features from Java projects mined from GitHub.
We used the ManySStuBs4J dataset \cite{karampatsis2020often}, which presents a classification of common single statement bug instances from 1000 Java projects on GitHub. We extended the original dataset by mining additional GitHub projects to identify even more bug instances from the newly added projects. The name of each classification is self-explanatory as it explains violations of certain rules that led to the bug, and in most cases, it also describes the fix of the potential bug. This includes patterns like \textit{Change Identifier Used} and \textit{Same Function More Args}. 
%(identifier appearing in the statement was replaced with another one)
%(an overloaded version of the function with more arguments was called). 

As a result of our study, we found 139 dynamic language features-related bug instances from the  mined projects.
%by mining additional GitHub projects aiming to identify even more bug instances from the newly added projects. 
%In total, we found 139 single statement bug instances that relate to dynamic features. %, with 32 of those bugs being reported in the project's issue tracking system. Most of those bugs are related to the use of reflection.%\todo[]{I don't quite understand this one. so we found 58 bug instances in total. What does the 32 refer to? Li: they are referring to issues created for that bugs.}. 
The majority of these bugs belong to three patterns: \textit{Wrong Function Name} (34\%), \textit{Same Function More Args} (18\%) and \textit{Change Identifier Used} (15\%). These findings can potentially provide guidance for automated program repair to detect and automatically patch dynamic language features related bugs. 

% we hope that they can inspire future work in this area. Bugs with single line change fixes grant an opportunity to easily identify dynamic feature-related bugs, understand their causes, and if these bugs present particular challenges for automatic repair.  

\section{Related Work}

 Bugs related to underdetermined specifications in the reflection API are discussed in Pontes et al. \cite{pontes2019}. These are bugs in JDK implementations, not applications or libraries. It seems unlikely that bugs of this nature (underdetermined API assumptions) in application code can be fixed with single statement changes.   
The study by Chen et al.  \cite{chen2018study} is an extensive study of Python bug fixes that have changes involving dynamic features.
Zhang et al. \cite{zhang2021} discuss a case where fixing a bug related to an underdetermined reflection API method (the order of the elements returned by \texttt{getDeclaredFields}) adds multiple lines to the code. Bug patterns involving Java streams are discussed in Khatchadourian et al. \cite{khatchadourian2020empirical}. Java exception handling bugs are explored in Ebert et al. \cite{ebert2015exploratory}. 
 
%\subsubsection{Simple bugs studies}
ManySStuBs4J \cite{karampatsis2020often} is a dataset of over 153k single statement bug fix changes mined from 1,000 popular open-source Java projects. We used this dataset to locate dynamic language feature bug fixes, but we extended the original dataset with 1,032  additional GitHub projects. Another similar dataset is CodeRep \cite{chen2018codrep}, which is a single-line changes dataset mined from different repositories, which has been used in program repair studies. Bugs2Fix \cite{tufano2019} is a data set of simple bugs used in program repair and bug-related studies. Unlike these two datasets, ManySStuBs4J focuses on fix templates and fixes that are at the statement level. 
% two other single statement bug datasets
%\subsubsection{Bug datasets}
%Defects4J \cite{just2014defects4j}, a well known dataset of reproducible bugs from real-world Java programs, and Bugs.jar \cite{saha2018bugs}, a larger dataset, are intended for and have been used in patch/repair, debugging and testing studies for Java programs. These datasets do not contain strictly single statement bug fix changes, whereas from the dataset in our study, each bug fix is due to a single line of code. Further, those datasets are not strictly restricted to the use of dynamic language features. 
% question here is how many bugs in defects4j are due to or involve reflection?

%\subsubsection{Language-feature related bugs}

%\todo[inline]{what features? how frequent? explain further}

% @Shawn thoughts on MOTIVATITION/METHODOLOGY
%What is the symptom: a reflection related exception? unexpected behaviour? is the behaviour captured in tests, does the fix have regression tests
% What causes the bug: not understanding the API, missing the protection of compile-time typechecking, changes that do not indicate it's an issue with reflection usage.

%I feel like i have done the background in the introduction. maybe we dont need the background, having a related works instead?
% @Shawn I think we may need to change the introduction based on an agreement on positioning the paper. my view is reflected in suggested changes to abstract

\section{Methodology}
\label{sec:methodology}

\begin{figure}[h]
  \centering
      \includegraphics[width=1\columnwidth]{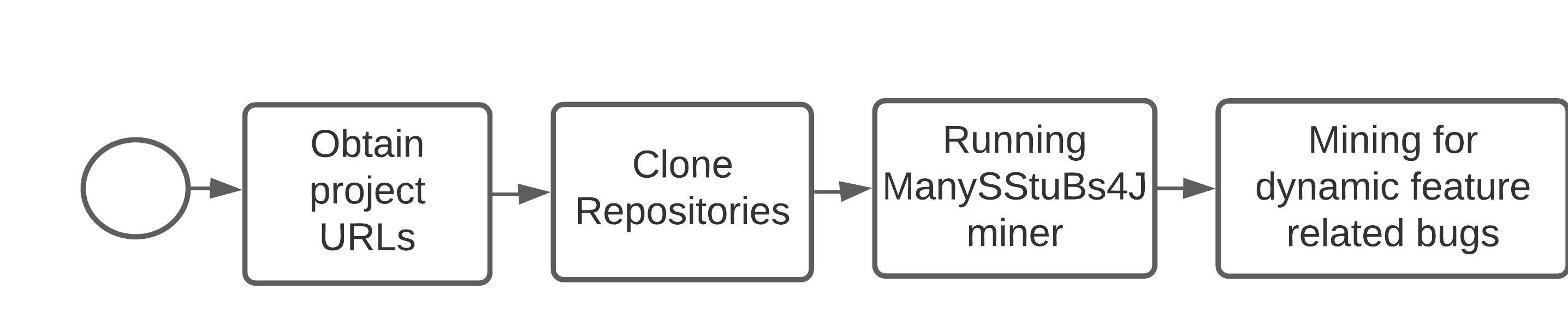}
  \caption{An overview of our mining and data extraction process}
  \label{fig:process}
\end{figure}
%We discuss the methodology used in our study in this section. 
Figure \ref{fig:process} provides an overview of the data collection process that we followed in this paper. As a first step, we gathered project URLs from (1) the ManySStuBs4J \cite{karampatsis2020often}, (2) additional projects from six organizations and communities hosted on GitHub (namely Google, %\footnote{\url{https://github.com/google} [accessed on 12,May 2021]}, 
Eclipse, %\footnote{\url{https://github.com/eclipse} [accessed on 12,May 2021]}, 
JetBrains, %\footnote{\url{https://github.com/JetBrains}}, 
Mozilla, %\footnote{\url{https://github.com/mozilla}}, 
Apache, %\footnote{\url{https://github.com/apache} [accessed on 12,May 2021]} 
and Spring.) %\footnote{\url{https://github.com/spring-projects} [accessed on 12,May 2021]}) in order to obtain more dynamic language feature related bugs. 
%\todo[]{why? isn't because we didn't find enough in the original dataset}. 
We identified those organizations and communities based on the following criteria: Firstly, they contain projects that are known for heavy use of dynamic language features, such as Spring. Secondly, they have a number of well-known Java projects, indicated by GitHub star counts. Thirdly,  the availability of an issue tracker, either hosted on GitHub or externally, such as on Bugzilla.
% \item all projects are currently active and being constantly updated.

%We extracted bug instances based on specific keywords that represent the use of dynamic language features in Java (we explain this in detail later in this section). 
%\todo[]{but you didn't say why we did so - how many we found in the original dataset? Li: we mentioned in the introduction. we found 51 unique bug instances in the original dataset, 7 in the additional dataset }

%Those six communities or organisations have a large number of projects so
%We harvested their main GitHub pages for URL's of the projects in order to clone the repositories. We use HTTP GET requests to retrieve the pages with the list of repositories and used an HTML parser %(Jsoup\footnote{\url{https://jsoup.org} [accessed on 12,May 2021]}) 
We first removed duplicate projects (i.e., projects that have already been included in the ManySStuBs4J dataset). We then cloned all remaining projects locally.\footnote{All projects were cloned on 12 Nov 2021, so the mining process is accomplished on the latest commit available on that date.}
%\todo{how many duplicates? how many projects have you obtained in total? Li: I did not record the total as I exclude them when parsing the page.}.
To extract single statement bugs from each repository, we used the same script 
%\todo{explain here how this works. does it mine issues in the repositories or fixes in the code as well? }
used in the original ManySStuBs4J study \footnote{\url{https://git.io/JnS32} %[accessed on 12,May 2021]
}.

%Next, we combined the data obtained from both datasets (the original ManySStuBs4J and the additional projects we identified), 
We developed a script to identify single statement bugs that are related to dynamic language features. The script searches for specific keywords at call sites in the parent commit of fix commits (where the bug occurs). These methods and keywords are based on a benchmark of dynamic feature usage patterns by Sui et al. \cite{sui2018soundness} and a list of the Java Reflection APIs from Landman et al. \cite{landman2017challenges}. We have excluded certain APIs from \cite{landman2017challenges}, such as logic operators (i.e., ==, !=) and call sites that can cause false positives, e.g. \texttt{toString()}, \texttt{set} and \texttt{get} as those keywords can be too generic. There are a total of 106 call sites selected for keyword matching. We acknowledge that this is not an exhaustive list of dynamic language feature methods and keywords in Java. To the best of our knowledge, there are no
studies that have comprehensively listed all the dynamic language feature methods and keywords in Java. Acquiring such a comprehensive list is outside the scope of this study.

%The methods listed in the table correspond to the APIs that allow customisation of the class and object life cycle (class loading/definition, object instantiation), introspection, customising execution semantics (dynamic proxies), and dynamic method invocation. 
% \todo[inline]{I think there's a validity issue here. This list of DLF methods is incomplete. For e.g., getField is not listed while getDeclaredField is. This doesn't seem consistent. Examples of other missing methods: getField\textbf{s}, getDeclaredField\textbf{s}.  Landman et al. have a more complete list of reflection DLF methods. Perhaps some are missing here because Sui et al. study DLF features  motivated by recall issues in SA tools due to dynamic object construction/method invocation and hence only those fewer methods? Moreover, the benchmark paper doesn't give a listing of methods and coverage of DLF (especially reflection methods) aren't exhaustive in the benchmark. Even if we list fewer methods here there's still inconsistency as proven by the get*Field case}

The dataset includes bug data from 2,024 Java projects, with a total of 249,089 bug instances being identified. 
%Those projects were selected based on their popularity index (based on number of forks and stars). The dataset contains a total of 153,652 bugs. 
From those projects, we mined a total of 104,337 single statement bugs (i.e., bugs that match the patterns identified in Karampatsis and Sutton \cite{karampatsis2020often}) searching for the use of specific dynamic language features. %We then mined an additional 1,032 Java projects from the six  organization/communities public GitHub repositories . % 1,908 before remove duplicates

 %Li yes, we may find additional dynamic features, those we found are based on ManySStuBs4 bug pattern.

 We provide the scripts used to extract the data together with our full results in a replication package\footnote{\url{https://https://github.com/li-sui/miningSStuBs}}.

\section{Results and Discussion}

%stats: additional dataset: 1032 projects, SStuBs: 13718(19DLF) total bugs: 33137(77DLF)
%ManySStuBs4J, 1000projects, SStuBs:63,923 (120DLF) total bugs: 153,652 (295 DLF)

%After filtering by keywords provided in Table \ref{tab:keyword},
By filtering each bug instance using the dynamic features keywords,
%\todo{so we did this by looking at each fix, Li: before fix}, 
we identified 1,916 dynamic feature related bugs. To verify the results, we developed a script to perform source code analysis on a limited scope (at class-level) to reduce potential false positives (e.g., call sites that match any of the keywords, but are not actually dynamic features\footnote{An example of keyword match that turns out not to be a dynamic feature is shown here \url{https://git.io/JnDyz}}). This analysis reduces the total number to 398 dynamic feature related bugs. At the end, two other authors conducted a manual validation by inspecting each bug instance to further check whether those bugs actually represent dynamic language feature related bugs (limited scope source code analysis may also cause false positives), and also to identify possible duplicates (e.g., same changes/fix but appear in multiple commits). %\todo[inline]{merged commits? not clear how this leads to duplicates. duplicates in what?}
% the commit \footnote{\url{https://github.com/junit-team/junit4/commit/b1068dc55c95d1000e07b71d3a9d9a2fd26bfff9}} is a merge for the commit \footnote{\url{https://github.com/junit-team/junit4/commit/eb307fa6f37ac8ec695b2f556e543969503418f5}} )
%We found 219 false positives and 27 duplicates in total. Those were removed from any further analysis. This has resulted in 139 bug instances that are classified to be related to the use of dynamic language features. 
%We then filter out instances that do not match the bug patterns provided in ManySStuBs4J,
%we identified 120 dynamic feature related bugs in ManySStuBs4J and another 19 bugs in the newly added projects. . We found 13 duplicates and 68 false positives. Those were removed from any further analysis. 
%which resulted in identifying 62 dynamic features single statement bugs. However, we found that four of these bug instances appear across multiple categories. For instance, one of the bugs that was found in Junit4\footnote{\url{https://git.io/JnDyb}} matched two reflection patterns: \texttt{Class.getMethod()} and \texttt{Method.invoke()}. 
We ended up with 139 unique single statement bugs in the datasets.
%(51 bugs that are attributed to dynamic language features in ManySStuBs4J and seven bugs from the additional projects that we analysed).

We note that the number of bugs found, 139, is relatively low compared to the total number of bug instances in the dataset. There are some possible explanations for the low occurrence of fixes involving direct changes to the use of dynamic language features. Firstly, those features are primarily used in code written to be reused (i.e., in upstream artefacts that are likely to be well-tested) which might be the case for some projects. Secondly, while the use of these features is often not safeguarded by the compiler, it is safeguarded by the usage context or a framework. For instance, consider data binding, where Java objects are mapped to some structured data representation (JSON, XML, etc.). When data is read, object state is initialized by invoking setters. However, these setters are not controlled by the programmer, but extracted from classes on-the-fly.
%So bug patterns like using the \textit{wrong method name} or the \textit{wrong number of parameters} can never occur. %Finally, it could be because we have not fully covered all dynamic features in our selected list of keywords.

Detailed statistics of the distribution of bugs across different dynamic language features (grouped by categories) is shown in Table \ref{tab:results-overview}. Note that the total number is not 139 as a single bug with multiple fixes can contain different dynamic call sites.

\begin{table}[h]
\centering
\caption{Distribution of single statement bugs across different dynamic features categories}
%\scriptsize
\label{tab:results-overview}
%\footnotesize

\begin{tabular}{l|r}
\hline
\textbf{Call site}       &  \multicolumn{1}{l}{\begin{tabular}[c]{@{}l@{}}\textbf{No. bugs}\end{tabular}} \\ \hline
java.lang.Class::get*               & 88                                                                            \\ 
java.lang.Class::forName               & 16                                                                             \\ 
java.lang.reflect.Method::invoke               & 13                                                                             \\ 
java.lang.Class::is*                  & 12                                                                               \\
java.lang.ClassLoader::*                 & 10                                                                              \\ 
java.io.ObjectInputStream::readObject                  & 3                                                                              \\ 
java.lang.Class::newInstance                 & 3                                                                              \\ 
java.util.ServiceLoader::load                & 2                                                                              \\
java.lang.reflect.Proxy::getInvocationHandler& 1                                                                              \\ \hline
\end{tabular}
\end{table}

To determine how the changes (fix commit - the commit that fixes the bug) on dynamic language features changes relate to the program behaviours, we specified them into three relationships: strong, intermediate and weak. (1) A strong relationship implies the fix also involves using a dynamic language features - it could be the same call site (change from \texttt{Method.invoke(this,null)} to \texttt{Method.invoke(this, "null")}) or different one (change from \texttt{getClass()} to \texttt{getDeclaredClass()}). (2) Intermediate indicates the changes are not directly on the dynamic language call site, but a dynamic language features still in used (e.g.,  \texttt{type.equals(number.\\getClass()))} to \texttt{type.isAssignableFrom(number.getClass())}).\\(3) A weak relationship indicates that there is a removal of dynamic language features or the changes has nothing to do with the dynamic language call site (e.g. a change on the logical operator\footnote{\url{https://git.io/J9GL6}} ). In summary, we identified 105 strong, 20 intermediate and 14 weak relationships.

The 139 single statement bugs are distributed across 79 projects.
We also checked whether those bugs were reported in the issue tracker system of the project so that we could read through the issue/comments to better understand the cause of the bug. We found 80 of the 139 bugs to have related issues in the issue tracking system. In most cases, with issues created for those bugs, we observed that there are more details about the nature of the bug, how it occurred and how it was fixed. For those 80 bugs, we analysed the discussion in the issue trackers in order to classify those bugs.

We mapped each bug into the bug patterns that were defined in \cite{karampatsis2020often}. Table \ref{tab:results-fixmethod} shows the distribution of dynamic feature-related single statement bugs we found across the different bugs patterns.
%In Table \ref{tab:results-fixmethod}, there are ?? bugs as three of these bugs match more than one pattern. For instance, a bug identified in Apache Felix\footnote{\url{https://git.io/J93K5}} matches the description of \textit{Same Function Wrong Caller} and \textit{Same Function More Args}.
The results also show that the majority of bugs belong to two patterns: \textit{Wrong Function Name} (34\%) and \textit{Same Function More Args} (18\%)
We discuss the top five bug patterns we found, with examples for each of these patterns below.
%Most of bugs have been resolved via either providing more arguments for overloaded methods (OVERLOAD\_METHOD\_MORE\_ARGS) or changing arguments (CHANGE\_IDENTIFIER). This indicates dynamic features are likely to be wrongly handled when passing arguments.

\begin{table}[h]
\centering
%\scriptsize
\caption{Single statement bug patterns}
%\footnotesize
\label{tab:results-fixmethod}
%\resizebox{0.80\linewidth}{!}{
\begin{tabular}{l|r}
\hline
\textbf{Pattern name}                           & \multicolumn{1}{l}{\textbf{No. of bugs}} \\ \hline
Wrong Function Name &47\\
Same Function More Args      & 25                          \\ 
Change Identifier Used       & 21                          \\ 
Same Function Wrong Caller & 19                          \\ 
Same Function Less Args    & 6                           \\ 
More Specific If                  & 7                          \\
Less Specific If                 & 8                           \\ 
Change Operand &4\\
Change Boolean Literal             & 1                           \\ 
Change Numeric Literal&1\\ \hline
\end{tabular}%}
\end{table}

\subsection{Wrong Function Name}\label{sec:wrongfunctionname}
This bug pattern describes a case where the method caller and the method call arguments are the same, but the name of the method that is different. In the context of reflection, this can appear as a method that has been used in place of another similar method. We observed 37/47 bug instances in this category, before fix and after fix share similar method name, and have the same return type. This could be caused by a confusion over API usage. An example is shown below, \texttt{java.lang.Class::getName} returns the name entity, but \texttt{getSimpleName} returns the name with \texttt{"[]"} appended.

\begin{table}[h]
  \footnotesize
  \label{tab:intellij}
  \begin{tabular}{rp{5cm}}
    before fix:  &  \texttt{con.getClass().getName()} \\
    after fix:  &  \texttt{con.getClass().getSimpleName()} \\
    bug pattern:   &  \textit{Wrong Function Name} \\
    fix commit: &  769753dac429f0448ecb56ee810e4f4a86f9fc9d \\
    commit message:   &  ResourceUtils.useCachesIfNecessary() not correct handle JNLP connections\\
    issue link &  \url{https://github.com/spring-projects/spring-framework/issues/14181}
  \end{tabular}
\end{table}

Another example of this category of bugs is shown in project M66B XPrivacy \footnote{\url{https://github.com/M66B/XPrivacy/pull/2203}}.% that is indicated by the commit message: \textit{``A fix for Groovy formatter tests....''}. 
% \begin{table}[h]
%   \footnotesize
%   \label{tab:intellij}
%   \begin{tabular}{rp{5cm}}
%     before fix:  &  \texttt{am.getClass().getDeclaredField("mContext");} \\
%     after fix:  &  \texttt{am.getClass().getField("mContext");} \\
%     bug pattern:   &  \textit{Wrong Function Name} \\
%     fix commit: &  200967dcf16cd3a72645c8668baec9ed83a0a7b7 \\
%     commit message:   &  Fixed on demand restricting on some Android Lollipop ROMs\\
%     issue link &  none
%   \end{tabular}
% \end{table}
The method \texttt{getDeclaredField}\footnote{\url{https://docs.oracle.com/javase/8/docs/api/java/lang/Class.html}} returns all declared fields within the class, whereas the method \texttt{getField} returns public member fields of the class. To fix this, the call site was changed from  \texttt{getDeclaredField} to \texttt{getField}, without changing the arguments. %Even though there is no related issue created for this bug, 
The fix implies that the intention of the call was to access public fields via reflection.

\subsection{Same Function More Args}\label{sec:samefunctionmoreargs}
The most common pattern we found is \textit{Same Function More Args}, with 25 bug instances found in 14 projects. We show an example of this bug from the popular JUnit framework. %\footnote{\url{https://junit.org/junit4/} [accessed on 4,June 2021]} .

\begin{table}[h]
  \footnotesize
  \label{tab:junit4}
  \begin{tabular}{rp{10cm}}
    before fix:  &  \texttt{Class.forName(name);}\\
    after fix:  &  \texttt{Class.forName(name,false,getClass().getClassLoader())} \\
                
    bug pattern:   &  \textit{Same Function More Args} \\
    fix commit: &  eb307fa6f37ac8ec695b2f556e543969503418f5 \\
    commit message:   &  Fix for \#359\\
    issue link &  \url{https://git.io/JnS3o}
  \end{tabular}
\end{table}

A bug describes an \texttt{ExceptionInInitializerError} that has been thrown when filtering tests by category, indicating there is an issue with class initialization. The solution is to add more arguments to the \texttt{Class::forName} method. The class \texttt{java.lang.Class} provides two overloaded methods to load a class: one  providing the fully qualified name for a class (i.e., \texttt{Class.name("name")}) and another that provides a detailed class loading scenario (i.e., \texttt{Class.name(\\"name", false, getClassLoader())}) by specifying on which class loader that the class will be loaded, and whether this class should be initialized or not.
%Class.name("name") is equivalent to full initialization : Class.name("name", true, getClassLoader())
In this case, a full initialization of a class is not required, therefore the second argument should be flagged as false.
%This is a small fix but could be hard for any static analyser/bug detection tool to pick it up as the bug happens at the runtime.

\subsection{Change Identifier Used}\label{sec:changeidentifier}
%\todo[inline]{use the same structure as the previous ones. Explain the pattern first then talk about the bug}
This pattern describes a bug where the fix involves replacing an identifier with another one. 
In total, we found 21 instances of this bug. An example of this bug from Apache Tomcat is shown below:

\begin{table}[h]
  \footnotesize
  \label{tab:spring}
  \begin{tabular}{rp{5cm}}
    before fix:  &  \texttt{getNext().invoke(this,method,args);} \\
    after fix:  &  \texttt{getNext().invoke(proxy,method,args);} \\
    bug pattern:   &  \textit{Change Identifier Used} \\
    fix commit: &  7446259923e31c0b79271c589873df551ba4a73c \\
    commit message:   &  Fix BZ52015.JdbcInterceptor passes not 'this' but 'proxy' to getNext().invoke.\\
    issue link &  \url{https://bz.apache.org/bugzilla/show_bug.cgi?format=multiple&id=52015}
  \end{tabular}
\end{table}

A bug was reported for class \texttt{org.apache.tomcat.jdbc.pool.\\JdbcInterceptor}. This abstract class is responsible for implementing JDBC interceptor. The bug is the result of not using the correct identifier in the method call. To fix this, the identifier argument (the target instance that the method is invoked upon) was changed from \texttt{this} (the current object) to \texttt{proxy} (a proxy object). 

%spring-projects/spring-framework  Method.invoke()  CHANGE_IDENTIFIER  Jms2MessageProducerInvocationHandler properly delegates to CachedMess…  Collaborator  type: bug https://github.com/spring-projects/spring-framework/issues/16566  https://github.com/spring-projects/spring-framework/commit/134e5a2aecf913bcc184d5d3ae845a13d5369078

\subsection{Same Function Wrong Caller}\label{sec:samefunctionwrongcaller}

In this bug, the function that is  executed is the same, but it is being called by a different caller. The fix is simple, and it involves changing the caller class in the call. An example of this bug from Apache Tomcat is shown below. 

\begin{table}[h]
  \footnotesize
  \label{tab:tomcat}
  \begin{tabular}{rp{5cm}}
    before fix:  &  \texttt{Class.forName(this.owner);} \\
    after fix:  &  \texttt{ReflectionUtil.forName(this.owner);} \\
    bug pattern:   &  \textit{Same Function Wrong Caller} \\
    fix commit: &  4461a96d8bb2c2f6132e5f370cf0d47ad18c7adb \\
    commit message:   &  bug 41797: CNFE/NPE thrown from function mapper when externalizing..\\
    issue link &  none
  \end{tabular}
\end{table}
 
The bug reported the use of a generic caller to a specific caller. The caller \texttt{java.lang.Class} has been replaced by its own implementation: \texttt{org.apache.el.util.ReflectionUtil}. Another bug\footnote{\url{https://git.io/JnS35}} found in DBeaver indicates a change of caller from \texttt{TIMESTAMP\\\_READ\_METHOD} to \texttt{TIMESTAMPTZ\_READ\_METHOD}. The method name and the rest of the arguments remain unchanged.
%\todo[inline]{as this is short, please add another example here but without a table  (just a reference to a commit in the repo}

\subsection{Same Function Less Args}\label{sec:samefunctionlessargs}
Unlike \textit{Same Function More Args}, this pattern requires fewer arguments to be included. We found a total of 6 bug instances that followed this pattern. We show an example of this category for a bug that was reported in the Calligraphy project \footnote{\url{https://git.io/JnS3H}}. The bug shown below is the opposite to the previous example, where a class should be fully initialized when calling \texttt{Class::forName}.
%\todo[inline]{explain this in more detail please. similar to the previous bug/category}

\begin{table}[h]
  \footnotesize
  \label{tab:Calligraphy}
  \begin{tabular}{rp{10cm}}
    before fix:  &  \texttt{Class.forName(} \\
                 &  \texttt{"android.support.v7.widget.Toolbar", false, null);} \\
    after fix:  &  \texttt{Class.forName(} \\
                & \texttt{"android.support.v7.widget.Toolbar");} \\
    bug pattern:   &  \textit{Same Function Less Args} \\
    fix commit: &  435319f273391c4f886f6908c2a7fc7c3dda9e13 \\
    commit message:   &  fixed check, actually need to instantiate it if it exists.\\
    issue link &  \url{https://git.io/JnS36}
  \end{tabular}
\end{table}

%\subsection{Summary}

\subsection{Further investigation of common causes of these bugs}
%\todo[inline]{ I moved the material we had in the discussion and combined it with the results section.I feel this reads much better this way, and it is still within the context of the results.}
We further investigated common usage patterns found in the detected bugs and reported the potential causes below. We present the reasons/causes for the top three patterns (i.e., \texttt{Class::forName}, \texttt{Method::invoke}, \texttt{Class::getDeclared|Class|Field|Method|\\Constructor|Name}). For each bug instance, we manually inspect the source code and the issue tracker to find out the cause and reason for the bug. This was done by one of the authors, and all instances were cross-validated by another co-author. Table \ref{tab:results-overview} shows an overview of a number of bugs for each call site and dynamic feature category. As explained in Section \ref{sec:methodology}, these categories were adopted from Landman et al. \cite{landman2017challenges} and Sui et al. \cite{sui2018soundness}. The raw results are available in our replication package\footnote{https://github.com/li-sui/miningSStuBs}.

\begin{enumerate}
\item{\texttt{Class::forName}}
\begin{itemize}
\item Passing an incorrect class name.
\item Need to delay initialization of static blocks when initialization is not appropriate in the loading context but at the time of class use. %Two examples are provided in Section \ref{sec:samefunctionmoreargs} and Section \ref{sec:samefunctionlessargs}.
\item Incorrect class loader specified.
%\item \texttt{Null} as class loader to prevent expensive class loader lookup from stack.
% TODO @Amjed: On above commented out line, not sure if I understand this one correctly. can you please rephrase/explain? @Shawn: commented out. TODO: recheck}
\item Using the wrong caller. Section \ref{sec:samefunctionwrongcaller} describes an example where a reflection wrapper is used.
\end{itemize}
\item{\texttt{Method::invoke}}
\begin{itemize}
\item Using the wrong caller.
\item Method invoked on wrong target object. %An example is provided in Section \ref{sec:changeidentifier}.
\end{itemize}
\item{\texttt{Class::getDeclared|Class|Field|Method|Constructor\\|Name}}
\begin{itemize}
%\item some fixes are due to failing tests.
%\item method not working for subclasses. \todo[]{plz clarify this one}
\item Trying to access a method/field that is inherited from superclasses.
%\item looking up overloaded methods.
\item Trying to access a method/field that is declared as a public or private member. %An example is provided in Section \ref{sec:wrongfunctionname}. %\todo{can't follow this one}
\item Using the wrong caller.  %-- in one case due to potential copy/pasted code.
\item Passing the incorrect arguments when accessing methods and constructors.
\item Using the wrong method. e.g. \texttt{Class::getName} vs \texttt{Class::\\getSimpleName}
\end{itemize}
\end{enumerate}
\section{Conclusion and Future Work}

% In this paper, we have studied \todo[inline]{discovered doesn't sound right} a class how the use of dynamic language features give rise to bugs in real-world programs, especially with single statement bugs. Surprisingly, only 58 unique bugs were found by mining a total of 97,060 single statement bugs.
% We want to point out the fact that those bugs are found within a single statement -- this provides an opportunity to easily identify call sites that are part of dynamic language feature. However, given the complexity of analysing such features, we strongly believe these findings can provide a guidance for program repair to detect and automatically patch dynamic language features related bugs. 

In this paper, we studied single statement bugs that are associated with the use of dynamic language features. We analysed data from 2,024 Java projects. From a total of 104,337 single statement bug instances, we identified only 139 unique bugs that relate to the use of dynamic language features. We attributed the relatively low number of bugs to how these features are used in practice (e.g.,  they are primarily  used  in  code written to be reused).  
Still, analysing these bugs provide an insight into this class of bugs in real-world programs. With the identification of these bugs, we hope
(1) this work helps practitioners to quantify the risk of using dynamic techniques over alternatives. (2) The work informs the developers of program analyses where to focus their efforts in order to produce tools with better recall.  (3) The work informs the developers and maintainers of reflective APIs about which APIs are particularly error-prone and may need refactoring or additional documentation.
%These findings can potentially provide guidance for program repair to detect and automatically patch dynamic language features related bugs. 
%We provide a  mini-dataset of dynamic features-related bugs that can be potentially used in evaluating program repair and static analysis tools. 
%A limitation of the approach is the bug fixes mined involve direct changes to statements that invoke these features. 

One of the limitations of this work is that we consider only a subset of dynamic language features, as we did not use a comprehensive list of all dynamic feature methods and keywords. To the best of our knowledge, such a list is not currently available. Expanding the list of methods and keywords would involve challenges in mining (e.g., \texttt{*.get} can yield false positives with the string matching approach, and we will need to statically analyse the code for precision). We expect the manual effort required for these tasks to be beyond the scope for this preliminary study and hence, it has been left for future work.

There are a number of open issues that will need to be investigated in the future, such as root cause analysis for these bugs and improving the precision/recall of mining. 
%A further analysis of such a bug, particularly with the use of reported stack traces, can help develop a better understanding of the root causes of these bugs (e.g., whether a bug is the result of a \texttt{ReflectiveOperation\\-Exception}).

%Moreover, reported stack traces can be used as input for static analysis tools \cite{sui2017use} to detect such bugs.
%A further step can be taken to develop plugins for IDEs to perform on-the-fly dataflow analysis for bug fixes, auto-complete and refactoring.

%We want to point out the fact that those bugs are found within a single statement -- this provides an opportunity to easily identify call sites that are part of dynamic language features. However, given the complexity of analysing such features, we strongly believe these findings can provide a guidance for program repair to detect and automatically patch dynamic language features related bugs. 

 %It is important for reflection-aware program repair as our findings contribute to understanding these bugs.  

% \begin{acks}
% \end{acks}

\bibliographystyle{ACM-Reference-Format}
\bibliography{main.bib}

\end{document}